\documentclass[%
 reprint,
 amsmath,amssymb,
 aps,
]{revtex4-1}

\usepackage{graphicx}
\usepackage{dcolumn}
\usepackage{bm}

\begin{document}

\preprint{APS/123-QED}

\title{Rotating spacetime: black-bounces and quantum deformed black hole}

\author{Zhaoyi Xu}
 \email{zyxu@gzu.edu.cn}
 \affiliation{%
College of Physics, Guizhou University, Guiyang 550025, China. \\
Key Laboratory of Particle Astrophysics,
Institute of High Energy Physics, Chinese Academy of Sciences,
Beijing 100049, China.
}%

\author{Meirong Tang}%
\affiliation{%
Yunnan Observatories, Chinese Academy of Sciences, 396 Yangfangwang, Guandu District, Kunming, 650216, China.\\
 University of Chinese Academy of Sciences, Beijing 100049, China.\\
 Key Laboratory for the Structure and Evolution of Celestial Objects, Chinese Academy of Sciences, 396 Yangfangwang, Guandu District, Kunming 650216, China.
}%

%
%

\date{\today}

\begin{abstract}
Recently, two kinds of deformed schwarzschild spacetime have been proposed, which are the black-bounces metric (\cite{2019JCAP...02..042S,2021PhRvD.103h4052L}) and quantum deformed black hole (BH) (\cite{2021arXiv210202471B}). In present work, we investigate the rotating spacetime of these deformed Schwarzschild metric. They are exact solutions to the Einstein$'$s field equation. We analyzed the properties of these rotating spacetimes, such as event horizon (EH), stationary limit surface (SIS), structure of singularity ring, energy condition (EC), etc., and found that these rotating spacetime have some novel properties.
\end{abstract}

\maketitle


\section{Introduction}
\label{intro}
In general relativity (GR), physicists understand the detailed properties of the gravitational field by solving the solutions to Einstein$'$s field equations (EE). 
However, because of the nonlinear characteristics of Einstein$'$s field equation, it is very difficult to solve it.
So far, physicists have only obtained some exact solutions in special cases, such as the Schwarzschild solution, Kerr solution and wormhole solution in vacuum.
These solutions have been studied extensively and deeply by physicists (\cite{1998bhp..book.....F}).
In recent years, the Laser Interferometer Gravitational-Wave Observatory (LIGO) gravitational wave measurements and the Black Hole Event Horizon Telescope (EHT) black hole shadow measurements have provided reliable evidence for the existence of black holes in the universe (\cite{2016PhRvL.116f1102A,2019ApJ...875L...1E}).
These observations further energized physicists, making the search for an exact solution to Einstein$'$s equations of the gravitational field all the more important.

In the work of \cite{2019JCAP...02..042S}, they proposed a new solution to a gravitational field equation that links a normal Schwarzschild black hole to a wormhole by simply introducing a parameter that can be written into the same mathematical expression. Through adjusting the range of values of the parameter, the space-time metric corresponding to Schwarzschild BH, normal black hole, unidirectional wormhole and MT wormhole can be obtained.
Since this work, people have done a lot of research on this black-bounce spacetime. For example, the properties of black hole/wormhole spacetime are studied by using gravitational wave echo signal (\cite{2020CQGra..37g5014C}), Huang and Yang studied the transformation of electrically charged Alice wormholes into BHs (\cite{2019PhRvD.100l4063H}), Bronnikov and Konoplya studied the influence of the black-bounce solution on the gravitational echo signal under the model world model (\cite{2020PhRvD.101f4004B}), Lobo et al. studied the dynamic properties of the black hole/wormhole transformation solution (\cite{2020PhRvD.101l4035L}), Nascimento et al. proposes to verify the properties of this transformation solution by gravitational lensing, and studies the absorption process of massless scalar waves under this transformation solution (\cite{2020arXiv200513096N,2020PhRvD.101l4009J}). Lobo and Rodrigues et al. extends the black-bounce solution to a more general case (\cite{2021PhRvD.103h4052L}). 
Recently, Jacopo Mazza, Edgardo Franzin and Stefano Liberati first generalized the Simpson-Visser(SV) spacetime to the Kerr case with Newman-Janis algorithm (NJA), and analyzed the properties of the exact solution (\cite{2021JCAP...04..082M}). Later in their work, they extended this situation to the Kerr-Newman spacetime (\cite{2021JCAP...08..036F}).

For this regular BH metric, there is another kind of solution that people pay much attention, that is quantum deformed black hole (\cite{1994NuPhB.429..153K,2016NuPhB.909..173A,2020arXiv200301333G}). Scientists can use it to link quantum gravitational effects to astronomical observations, such as gravitational lensing and black hole shadows. However, just as one has only the spherically symmetric metric, for the very important metric of a rotating BH, the exact solution has not been obtained.

However, these studies only focus on the exact solutions of spherical symmetry. As we know, for BHs or wormholes, the rotation case will be very interesting, so the study of the black-bounce solution and quantum deformed BH of rotation case will be of great significance. On the other hand, Newman-Janis algorithm (NJA) has long proposed that the rotation metric can be obtained through a series of complex transformations, in which the key is to transform the gravitational field equation into a set of partial differential equations (\cite{1965JMP.....6..915N,2014EPJC...74.2865A,2016EPJC...76....7A}). In this work, we will use the NJA to derive the exact solutions in the rotation case, and two special cases, namely the black-bounce and quantum deformed BH, are discussed and analyzed.

The outline of this work is as follows. In Sec \ref{emtc}, we introduce the deformed Schwarzschild solution, they include the black-bounce spacetime and quantum deformed BH. In Sec \ref{deei}, we derived the analytical form of deformed Kerr BH. In Sec \ref{pro}, we analysis the properties of these deformed Kerr BH metric. The summary are present in \ref{sum}.

\section{The regular Schwarzschild-like spacetime} 
\label{emtc}
\subsection{Black-bounce spacetime}
The regular spacetime metric is currently a hot topic of discussion in GR and BH physics, which gives possible distortions of the Schwarzschild metric. The black-bounce spacetime is described here. The black-bounce spacetime is the regular spacetime that connects the Schwarzschild BH to the traversable wormhole (\cite{2021PhRvD.103h4052L}). The general black-bounce spacetime is given by 
\begin{equation}
ds^{2}=-f(r)dt^{2}+\dfrac{1}{g(r)}dr^{2}
+h(r)(d\theta^{2}+sin^{2}\theta d\phi^{2}),
\label{SP1}
\end{equation}
where
\begin{equation}
f(r)=g(r)=1-\dfrac{2\mathcal{M}(r)}{\sqrt{h(r)}},$$$$
\mathcal{M}(r)=\dfrac{M\sqrt{h(r)}r^{k}}{(r^{2n}+m^{2n})^{\frac{k+1}{2n}}},$$$$
h(r)=r^{2}+m^{2},
\label{SP2}
\end{equation}
$M$ is the mass of the BH as the metric degrades to the Schwarzschild BH, $m$ is the non-negative parameter, which determine the type of metric (BH or traversable wormhole). $k$ and $n$ are positive integers, which equal to $0,1,2,3.....$. By valuing the model parameters, the spacetime metric can degenerate into the following types:

$\mathbf{A}$. Simpson-Visser model (\cite{2019JCAP...02..042S}), $k=0$ and $n=1$. For this case, the metric coefficients simplify to 
\begin{equation}
f(r)=g(r)=1-\dfrac{2M}{\sqrt{r^{2}+m^{2}}},$$$$
h(r)=r^{2}+m^{2},
\label{SP3}
\end{equation}
This space-time metric interpolates between a Schwarzschild BH and a traversable wormhole. For different value of parameter $m$, the space-time metric reduce to different space-time metric. Specifically, for $m=0$, spacetime metric is a Schwarzschild BH; for $0<m<2M$, spacetime metric is a regular BH; for $m=2M$, spacetime metric is a one-way traversable wormhole; for $m>2M$, spacetime metric is a Morris-Thorne traversable wormhole. 

$\mathbf{B}$. regular Bardeen-like BH model (with change $m$ to $q$, $\Sigma^{2}\rightarrow r^{2}$) (\cite{2021PhRvD.103h4052L}), $k=2$ and $n=1$. For this case, the metric coefficients simplify to 
\begin{equation}
f(r)=g(r)=1-\dfrac{2Mr^{2}}{(r^{2}+m^{2})^{\frac{3}{2}}},$$$$
h(r)=r^{2}+m^{2},
\label{SP4}
\end{equation}

The metric changes in various forms with different parameter values ($m$, $k$ and $n$), which makes the black-bounce spacetime contain a large number of space-time metric. Refer to \cite{2021PhRvD.103h4052L} for a discussion.

\subsection{Quantum deformed black hole}
Quantum deformed BHs due to quantum fluctuations in GR are introduced here. The quantum deformed BHs are caused by quantum fluctuations in two-dimensional dilaton gravity (\cite{1994NuPhB.429..153K,1992NuPhB.382..259R,2021arXiv210202471B}), the action is 
\begin{equation}
S=-\dfrac{1}{8}\int dz^{2}\sqrt{g}\big(r^{2}R^{(2)}+\dfrac{2}{\kappa}U(r)-2(\bigtriangledown r)^{2}\big),
\label{SP10}
\end{equation}
where $R^{(2)}$ is the Ricci scalar, $U(r)$ is the dilaton potential and $\kappa$ is the constant. By variating the action, the corresponding field equation can be obtained, and the space-time metric can be expressed as follows
\begin{equation}
ds^{2}=-f(r)dt^{2}+\dfrac{1}{g(r)}dr^{2}+r^{2}(d\theta^{2}+\sin^{2}\theta d\phi^{2}),$$$$
f(r)=g(r)=-\dfrac{2M}{r}+\dfrac{1}{r}\int^{r}U(\rho)d\rho.
\label{SP11}
\end{equation}
Based on this theory, Berry-Simpson-Visser (\cite{2021arXiv210202471B}) proposed a more general quantum deformed BH, which space-time metric is
\begin{equation}
f(r)=g(r)=-\dfrac{2M}{r}+(1-\dfrac{A^{2}}{r^{2}})^{\frac{n}{2}},
\label{SP19}
\end{equation}
where $n=0,1,3,5.....$, $r\geqslant A$ and $0<A<\infty$. Because of the different values of $n$, the spacetime metric can be classified as follows (\cite{2021arXiv210202471B}):

$\mathbf{A}$. Schwarzschild BH, $n=0$. The metric coefficients simplify to 
\begin{equation}
f(r)=g(r)=1-\dfrac{2M}{r},
\label{SP20}
\end{equation}

$\mathbf{B}$. Metric-regular, $n\geqslant 1$. 

$\mathbf{C}$. Christoffel-symbol-regular, $n\geqslant 3$.

$\mathbf{D}$. Curvature-regular, $n\geqslant 5$.

\section{Deformed Kerr spacetime}
\label{deei}
From spherical symmetry space-time metric to rotational symmetry space-time metric. There are many approaches to solve the Einstein$'$s field equation. Here, we use the NJ algorithm to study the exact solution of Einstein$'$s field equation with rotational case (\cite{1965JMP.....6..915N,2014EPJC...74.2865A,2016EPJC...76....7A}). The NJ algorithm main includes the complex transform and system of partial differential equations(PDE). 

According to the NJ algorithm, we need to transform the line element of spacetime to advanced null coordinates (ANC) $(u,r,\theta,\phi)$, and the transform equation is
\begin{equation}
du=dt-\dfrac{1}{f(r)g(r)}dr,
\label{NJA1}
\end{equation}
from the metric coefficients of black-bounce spacetime and quantum deformed BH, the metric coefficients $f(r)=g(r)=1-\dfrac{2M}{\sqrt{r^{2}+m^{2}}}\dfrac{r^{k}\sqrt{r^{2}+m^{2}}}{(r^{2n}+m^{2n})^{\frac{k+1}{2n}}}$ and $(1-\dfrac{A^{2}}{r^{2}})^{\frac{n}{2}}-\dfrac{2M}{r}$, then the equ. (\ref{NJA1}) reduce to
\begin{equation}
du=\left\{
             \begin{array}{lr}
             dt-\dfrac{\Big(r^{2n}+m^{2n}\Big)^{\frac{k+1}{n}}}{\Big((r^{2n}+m^{2n})^{\frac{k+1}{n}}-2Mr^{k}\Big)^{2}}dr, \\ \: \: \: \: \: \: \: \: \: \:{\rm for \, black \,-bounce \, spacetime;}  \\

             dt-\dfrac{1}{((1-\dfrac{A^{2}}{r^{2}})^{\frac{n}{2}}-\dfrac{2M}{r})^{2}}dr, \: \: \: \: \: {\rm for \, quantum \, BH.} \\ 
             \end{array}
\right.
\label{NJA2}
\end{equation}

In the advanced null frame, the corresponding inverse of spacetime metric is given by
\begin{equation}
g^{\mu\nu}=-l^{\mu}n^{\nu}-l^{\nu}n^{\mu}+m^{\mu}\overline{m}^{\nu}+m^{\nu}\overline{m}^{\mu},
\label{NJA3}
\end{equation}
where the base vector are
\begin{equation}
l^{\mu}=\delta^{\mu}_{r},
\label{NJA41}
\end{equation}

\begin{equation}
n^{\mu}=\left\{
             \begin{array}{lr}
             \delta^{\mu}_{\mu}-\dfrac{1}{2}\Big(1-\dfrac{2M}{\sqrt{r^{2}+m^{2}}}\dfrac{r^{k}\sqrt{r^{2}+m^{2}}}{(r^{2n}+m^{2n})^{\frac{k+1}{2n}}}\Big)\delta^{\mu}_{r}, \\ \: \: \: \: \: \: \: \: \: \:{\rm for \, black \,-bounce \, spacetime;}  \\

             \delta^{\mu}_{\mu}-\dfrac{1}{2}\Big((1-\dfrac{A^{2}}{r^{2}})^{\frac{n}{2}}-\dfrac{2M}{r}\Big)\delta^{\mu}_{r}, \: \: \: \: \: {\rm for\, quantum \, BH.} \\
             \end{array}
\right.
\label{NJA42}
\end{equation}

\begin{equation}
m^{\mu}=\left\{
             \begin{array}{lr}
             \dfrac{1}{\sqrt{2(r^{2}+m^{2})}}\delta^{\mu}_{\theta}+\dfrac{i}{\sqrt{2(r^{2}+m^{2})}\sin\theta}\delta^{\mu}_{\phi}, \\ \: \: \: \: \: \: \: \: \: \: {\rm for \,  black \,-bounce \, spacetime;}  \\

             \dfrac{1}{2r}\delta^{\mu}_{\theta}+\dfrac{i}{2r\sin\theta}\delta^{\mu}_{\phi}, \: \: \: \: \: {\rm for\, quantum \, BH.} \\ 
             \end{array}
\right.
\label{NJA43}
\end{equation}

\begin{equation}
\overline{m}^{\mu}=
\left\{
             \begin{array}{lr}
             \dfrac{1}{\sqrt{2(r^{2}+m^{2})}}\delta^{\mu}_{\theta}-\dfrac{i}{\sqrt{2(r^{2}+m^{2})}\sin\theta}\delta^{\mu}_{\phi},   \\  \: \: \: \: \: \: \: \: \: \:{\rm for \, black \,-bounce \, spacetime;}  \\

             \dfrac{1}{2r}\delta^{\mu}_{\theta}-\dfrac{i}{2r\sin\theta}\delta^{\mu}_{\phi}, \: \: \: \: \: {\rm for\, quantum \, BH.} \\ 
             \end{array}
\right.
\label{NJA44}
\end{equation}

From the advanced null frame, these base vectors satisfy orthogonality condition $l_{\mu}l^{\mu}=n_{\mu}n^{\mu}=m_{\mu}m^{\mu}=0$, $l_{\mu}n^{\mu}=-m_{\mu}\overline{m}^{\mu}=1$ and $l_{\mu}m^{\mu}=n_{\mu}m^{\mu}=0$. If we want to obtain the rotational space-time metric, we should need a complex transformation. In NJ algorithm, the transformation is
\begin{equation}
u\longrightarrow u-ia\cos\theta,$$$$
r\longrightarrow r+ia\cos\theta.
\label{NJA5}
\end{equation}
For this case, the corresponding metric coefficients should change to higher dimensions. Specifically, there are $f(r)\longrightarrow F(r,\theta,a)$, $g(r)\longrightarrow G(r,\theta,a)$ and $h(r)\longrightarrow \Psi(r,\theta,a)$ (here $h(r)=r^{2}+m^{2}$). Based on these changes, the new base vectors can be written as
\begin{equation}
l^{\mu}=\delta^{\mu}_{r},$$$$
n^{\mu}=\sqrt{\dfrac{G}{F}}\delta^{\mu}_{\mu}-\dfrac{F}{2}\delta^{\mu}_{r},$$$$
m^{\mu}=\dfrac{1}{\sqrt{2\Psi}}(\delta^{\mu}_{\theta}+ia \sin\theta(\delta^{\mu}_{\mu}-\delta^{\mu}_{r})+\dfrac{i}{\sin\theta}\delta^{\mu}_{\phi}),$$$$
\overline{m}^{\mu}=\dfrac{1}{\sqrt{2\Psi}}(\delta^{\mu}_{\theta}-ia \sin\theta(\delta^{\mu}_{\mu}-\delta^{\mu}_{r})-\dfrac{i}{\sin\theta}\delta^{\mu}_{\phi}).
\label{NJA6}
\end{equation}
Through these basis vectors, we can calculate the corresponding inverse metric component $g^{\mu\nu}$, which is given by
\begin{equation}
g^{uu}=\dfrac{a^{2}\sin^{2}\theta}{\Psi},~~~~~~g^{\theta\theta}=\dfrac{1}{\Psi},$$$$
g^{ur}=g^{ru}=\sqrt{\dfrac{G}{F}}-\dfrac{a^{2}\sin^{2}\theta}{\Psi},$$$$
g^{\phi\phi}=\dfrac{1}{\Psi\sin^{2}\theta},~~~~~~
g^{u\phi}=g^{\phi u}=\dfrac{a}{\Psi},$$$$
g^{r\phi}=g^{\phi r}=\dfrac{a}{\Psi}, ~~~~~~g^{rr}=G+\dfrac{a^{2}\sin^{2}\theta}{\Psi}.
\label{NJA7}
\end{equation}
From these inverse metric component, we obtain the Kerr-like space-time metric of black hole/wormhole transition in Eddington-Finkelstein coordinates (EFC), and they are
\begin{equation}
ds^{2}=-Fdu^{2}+2\sqrt{\dfrac{F}{G}}dudr+2a\sin^{2}\theta \big(\sqrt{\dfrac{F}{G}}+F \big)dud\phi $$$$-2a\sin^{2}\theta\sqrt{\dfrac{F}{G}}drd\phi+\Psi d\theta^{2}
-\sin^{2}\theta \big[-\Psi+$$$$ 
a^{2}\sin^{2}\theta \big(2\sqrt{\dfrac{F}{G}}+F \big) \big]d\phi^{2}.
\label{NJA8}
\end{equation}
Let$'$s define a new function $k(r)=h(r)\sqrt{f(r)}/\sqrt{g(r)}=(r^{2}+m^{2})\sqrt{f(r)}/\sqrt{g(r)}$ (for black-bounce spacetime) and $r^{2}\sqrt{f(r)}/\sqrt{g(r)}$ (for quantum deformed BH), we can convert space-time metric from EFC to Boyer-Lindquist coordinates (BLC) by the following relation
\begin{equation}
du=dt-\dfrac{k+a^{2}}{h(r)f(r)+a^{2}}dr,$$$$
d\phi=d\phi-\dfrac{a}{h(r)f(r)+a^{2}}dr.
\label{NJA9}
\end{equation}
According to the previous results, the functions $F(r,\theta)$ and $G(r,\theta)$ are
\begin{equation}
F(r,\theta)=-\dfrac{h(r)f(r)+a^{2}\cos^{2}\theta}{k(r)+a^{2}\cos^{2}\theta}\Psi,$$$$
G(r,\theta)=-\dfrac{h(r)f(r)+a^{2}\cos^{2}\theta}{\Psi}.
\label{NJA9}
\end{equation}
On the other hand, we make the following agreement, $\Sigma^{2}=k(r)+a^{2}\cos^{2}\theta$,  $2\overline{f}=k(r)-h(r)f(r)$, $\Delta(r)=h(r)f(r)+a^{2}$ and $A=(k(r)+a^{2})^{2}-a^{2}\Delta \sin^{2}\theta$. The spacetime metric in BLC is
\begin{equation}
ds^{2}=-\dfrac{\Psi}{\Sigma^{2}}\big(1-\dfrac{2\overline{f}}{\Sigma^{2}}\big)dt^{2}+\dfrac{\Psi}{\Delta}dr^{2}-\dfrac{4a\overline{f}\sin^{2}\theta \Psi}{\Sigma^{4}}dtd\phi+$$$$
\Psi d\theta^{2}+\dfrac{\Psi A \sin^{2}\theta}{\Sigma^{4}}d\phi^{2}.
\label{NJA10}
\end{equation}
From the spacetime metric (\ref{NJA10}), we don$'$t know the exact form of function $\Psi$. But the spacetime metric (\ref{NJA10}) has to satisfied two conditions, the rotational symmetric $G_{r\theta}=0$ and Einstein$'$s field equation $G_{\mu\nu}=8\pi T_{\mu\nu}$. At the same times, we set $y=\cos\theta$, $\Psi_{,ry^{2}}=\partial^{2}\Psi/\partial r \partial y^{2}$ and $k_{,r}=\partial k(r)/\partial r$.  Let$’$s plug spacetime metric (\ref{NJA10}) into above two equations, and these conditions become
\begin{equation}
(k+a^{2}y^{2})^{2}(3\Psi_{,r}\Psi_{,y^{2}}-2\Psi\Psi_{,ry^{2}})=3a^{2}k_{,r}\Psi^{2},
\label{NJA11}
\end{equation}
and 
\begin{equation}
\Psi[k^{2}_{,r}+k(2-k_{,rr})-a^{2}y^{2}(2+k_{,rr})]+(k+a^{2}y^{2})(4y^{2}\Psi_{,y^{2}}$$$$
-k_{,r}\Psi_{,r})=0.
\label{NJA12}
\end{equation}
For the spacetime metric of black-bounce, $f(r)=g(r)$, $h(r)=r^{2}+m^{2}$, therefore $k(r)=h(r)\sqrt{f(r)}/\sqrt{g(r)}=r^{2}+m^{2}$. From Eqs. (\ref{NJA11}) and Eqs. (\ref{NJA12}), we obtain the exact form of function $\Psi$, the result is $\Psi=r^{2}+m^{2}+a^{2}\cos^{2}\theta$. For the spacetime metric of quantum deformed BH, $f(r)=g(r)$, $h(r)=r^{2}$, therefore $k(r)=h(r)\sqrt{f(r)}/\sqrt{g(r)}=r^{2}$.  Therefore, the rotating spacetime metric of black -bounce spacetime and quantum deformed BH are:

$\mathbf{A}$. Rotating spacetime of black-bounce

\begin{equation}
ds^{2}=-\Big(1-\dfrac{2M(r^{2}+m^{2})r^{k}}{\Big(r^{2n}+m^{2n}\Big)^{\frac{k+1}{2n}}\Sigma^{2}}\Big)dt^{2}+\dfrac{\Sigma^{2}}{\Delta}dr^{2}$$$$
-\dfrac{4a\sin^{2}\theta}{\Sigma^{2}}\dfrac{M(r^{2}+m^{2})r^{k}}{\Big(r^{2n}+m^{2n}\Big)^{\frac{k+1}{2n}}}d\phi dt+\Sigma^{2}d\theta^{2}+$$$$
\dfrac{\sin^{2}\theta}{\Sigma^{2}}\Big((r^{2}+m^{2}+a^{2})^{2}-a^{2}\Delta\sin^{2}\theta\Big)d\phi^{2},
\label{Kerr-like1}
\end{equation}
where $\Sigma^{2}=r^{2}+m^{2}+a^{2}\cos^{2}\theta$ and $\Delta=h(r)f(r)+a^{2}=r^{2}+m^{2}-\dfrac{2Mr^{k}(r^{2}+m^{2})}{\Big(r^{2n}+m^{2n}\Big)^{\frac{k+1}{2n}}}+a^{2}$. When the non-negative parameter satisfies $k=0$, $n=1$ and BH spin $a=0$, the Kerr-like space-time metric of black-bounce degenerates into SV spacetime (\cite{2019JCAP...02..042S}).
If the non-negative parameter $m=k=0$ and $n=1$, the Kerr-like space-time metric of black-bounce degenerate into a Kerr BH. If the BH spin $a=0$, this Kerr-like space-time metric of black-bounce degenerate into a general spherical symmetry black-bounce spacetime (\cite{2021PhRvD.103h4052L}). As the parameter $m$ goes from $0$ to infinity, then the space-time metric becomes Kerr BH, Kerr-like BH, one-way rotating traversable wormhole and rotating Morris-Thorne traversable wormhole. As the values of parameters $n$ and $k$ increase, the rotation metric represents many BH and wormhole metrics. 

A special case of this rotating space-time is discussed. When $k=0$, $n=1$, the rotating space-time degrades into the rotating SV space-time, which has been studied in detail by Jacopo Mazza, Edgardo Franzin and Stefano Liberati (\cite{2021JCAP...04..082M}). This is also the most important special case of rotating spacetime (\ref{Kerr-like1}). Like SV spacetime, this rotating space-time contains both the rotating BH space-time and the rotating wormhole space-time.

$\mathbf{B}$. Rotating spacetime of quantum deformed BH

\begin{equation}
ds^{2}=-\Big(1-\dfrac{r^{2}+2Mr-r^{2}\Big(1-\dfrac{A^{2}}{r^{2}}\Big)^{\frac{n}{2}}}{\Sigma^{2}}\Big)dt^{2}+\dfrac{\Sigma^{2}}{\Delta}dr^{2}-$$$$
\dfrac{2a\sin^{2}\theta}{\Sigma^{2}}\Big(r^{2}+2Mr-r^{2}\Big(1-\dfrac{A^{2}}{r^{2}}\Big)^{\frac{n}{2}}\Big)d\phi dt
+\Sigma^{2}d\theta^{2}+$$$$
\dfrac{\sin^{2}\theta}{\Sigma^{2}}\Big((r^{2}+a^{2})^{2}-a^{2}\Delta\sin^{2}\theta\Big)d\phi^{2},
\label{Kerr-like2}
\end{equation}
where $\Sigma^{2}=r^{2}+a^{2}\cos^{2}\theta$ and $\Delta=h(r)f(r)+a^{2}=r^{2}\Big(1-\dfrac{A^{2}}{r^{2}}\Big)^{\frac{n}{2}}-2Mr+a^{2}$.
When the parameter $A=0$, the Kerr-like space-time metric of quantum deformed BH degenerate into a Kerr BH. If the BH spin $a=0$, this Kerr-like space-time metric of quantum deformed BH degenerate into a general spherical symmetry quantum deformed BH (\cite{2021arXiv210202471B}). Since the parameter $A$ represents the quantum correction of the BH, the value of $A$ should be very small, so the quantum correction Kerr BH is very close to the Kerr BH. On the other hand, As the values of parameters $n$, the Kerr-like BH metric represents many BHs.

\section{Properties of deformed Kerr spacetime}
\label{pro}

\subsection{Black-bounce spacetime}

$\mathbf{A}$. Rotating spacetime of black-bounce

If the non-negative parameter $m$ satisfies $0\leq m<m_{c}$ (where $m_{c}$ is the threshold at which the metric changes from a BH to a wormhole, We'll talk about that later), the spacetime metric (\ref{Kerr-like1}) degenerate into a regular Kerr BH.  
In the classic GR and BH physics, the properties of BH are determined by line element (\ref{Kerr-like1}). Here we consider horizon structure, stationary limit surfaces and singularity structure. 

Firstly, the horizon structure of BH is determined by the equation $\Delta=0$, then the event horizon (EH) $r_{+}$ and Cauchy horizon (CH) $r_{-}$ are 
\begin{equation}
r_{\pm}^{2}+m^{2}-\dfrac{2Mr_{\pm}^{k}(r_{\pm}^{2}+m^{2})}{\Big(r_{\pm}^{2n}+m^{2n}\Big)^{\frac{k+1}{2n}}}+a^{2}=0
\label{Pro-BB0}
\end{equation}
The appearance of $m$ makes the radius of the EH of regular BHs more interesting. For example, if consider SV model ($k=0$ and $n=1$), the critical parameter $m_{c}$ will be inconsistent with the spin of different BHs, and $m_{c}$ can be found by EH formula
\begin{equation}
m_{c}=\sqrt{2M^{2}-a^{2}+2M\sqrt{M^{2}-a^{2}}}.
\label{Pro-BB2}
\end{equation}
From the expression for $m_{c}$, when we consider that regular Kerr BH reduces to spherical symmetry case ($(a=0)$), the critical parameter $m_{c}=2M$. when the regular Kerr BH close to extreme BH ($(a=M)$), the critical parameter $m_{c}=\sqrt{2}M$. Therefore, the critical parameter $\sqrt{2}M\leqslant m_{c}\leqslant2M$.

Secondly, the stationary limit surfaces (SLS) of BH are determined by following condition
\begin{equation}
1-\dfrac{2M(r_{\rm{SLS\pm}}^{2}+m^{2})r_{\rm{SLS\pm}}^{k}}{\Big(r_{\rm{SLS\pm}}^{2n}+m^{2n}\Big)^{\frac{k+1}{2n}}(r_{\rm{SLS\pm}}^{2}+m^{2}+a^{2}\cos^{2}\theta)}=0.
\label{Pro-BB10}
\end{equation}
For the SV model ($k=0$ and $n=1$), the two SLS are 
\begin{equation}
r_{\rm{SLS\pm}}=\sqrt{2M^{2}-a^{2}\cos^{2}\theta-m^{2}\pm 2M\sqrt{M^{2}-a^{2}\cos^{2}\theta}}. 
\label{Pro-BB11}
\end{equation}
Through analysis, it can be found that the parameter $m$ can reduce $r_{\rm{SLS\pm}}$. If $m$ keeps increasing, then $r_{\rm{SLS+}}$ would keep decreasing until it is zero, at which point the BH event horizon and SIS would disappear, and the regular BH metric would transition to the wormhole. 

Thirdly, for a rotational regular BH, it has ring singularity, and the radius of the ring is $a$. For regular BH (\ref{Kerr-like1}), we find that Kretsmann scalar $R=R^{\mu\nu\rho\delta}R_{\mu\nu\rho\delta}=Z/\Sigma^{12}$, where $Z$ is a polynomial function. Then, we find that there have singularity at $r=\sqrt{a^{2}-m^{2}}$ and $\theta=\pi/2$. Therefore, the singularity reduce to point with $m=a$, this is a interest property. 

$\mathbf{B}$. Rotating spacetime of wormhole

If the non-negative parameter $m$ satisfy $m>m_{c}$ and $m=m_{c}$, the spacetime metric (\ref{Kerr-like1}) degenerate into a rotational Morris-Thorne traversable wormhole and a rotational one-way traversable wormhole, respectively. As we know from the previous analysis, the EH, SIS and the singularity ring disappear, so the space-time metric (\ref{Kerr-like1}) would have no singularity, and the whole space-time would be connected as one. For rotating wormhole spacetime, the metric obtained by the NJ algorithm can be written as follows (\cite{2016EPJC...76....7A})
\begin{equation}
ds^{2}=-\dfrac{\Sigma^{2}\Delta}{\widetilde{\Sigma}}dt^{2}+\dfrac{\Sigma^{2}\widetilde{\Sigma}}{\Delta}\dfrac{dr^{2}}{1-\dfrac{b(r)}{r}}+\Sigma^{2}d\theta^{2}+\dfrac{\widetilde{\Sigma}\sin^{2}\theta}{\Sigma^{2}}(d\phi-\omega dt)^{2}, 
\label{Pro-BB20}
\end{equation}
where $\widetilde{\Sigma}=(r^{2}+m^{2}+a^{2})^{2}-a^{2}\Delta\sin^{2}\theta$ and $\omega=2af(r)/\widetilde{\Sigma}$. Some preliminary analyses are as follows:

Firstly, by calculating the Kretsmann scalar $R$, we know that the metric (4.5) has no singularity when $m>m_{c}$, which is a direct feature of the wormhole spacetime. 

Secondly, For the rotating wormhole metric, the shape factor is a very important physical quantity. In order to obtain the shape function of the rotating wormhole, refer to the Morris-Thorne metric form (equation 11 in \cite{2016EPJC...76....7A}) , and the following conditions should be met between the metric coefficients
\begin{equation}
1-\dfrac{B(r,\theta)}{r}=\dfrac{\Delta}{\Sigma^{2}\widetilde{\Sigma}}\Big(1-\dfrac{b(r)}{r}\Big), 
\label{Pro-BB21}
\end{equation}
when consider the geometry of the throat ($r=r_{0}$), the $B(r_{0},\theta_{0})=r_{0}$. At this time, the geometry near the throat of the rotating wormhole is a complex hypersurface, rather than an a spherical surface. 

Thirdly, If the derivative of equation (4.6) in the $r$ direction is taken, the flare-out condition can be obtained, which is expressed as
\begin{equation}
\dfrac{1}{r^{2}}\Big(B(r,\theta)-r\dfrac{\partial B(r,\theta)}{\partial r}\Big)=\dfrac{\Delta}{r^{2}\Sigma^{2}\widetilde{\Sigma}}\Big(b(r)-r\dfrac{db(r)}{dr}\Big)$$$$
+\dfrac{\partial}{\partial r}\Big(\dfrac{\Delta}{\Sigma^{2}\widetilde{\Sigma}}\Big)\Big(1-\dfrac{b(r)}{r}\Big), 
\label{Pro-BB22}
\end{equation}
when $r\rightarrow r_{0}$, the $1-\dfrac{b(r)}{r}\rightarrow 0$, therefore, the flare-out condition make $B(r,\theta)$ satisfied 
\begin{equation}
r\dfrac{\partial B(r,\theta)}{\partial r}<B(r,\theta), 
\label{Pro-BB23}
\end{equation}
It$'$s the same condition as a spherically symmetric wormhole. Due to the $B(r_{0},\theta_{0})=r_{0}$, the flare-out condition reduce to $\dfrac{\partial B(r,\theta)}{\partial r}\mid_{r_{0},\theta_{0}}\leq 1$. 

Fourthly, the rotating wormhole metric satisfies the asymptotic condition, that is, when $r\rightarrow \infty$, the spacetime metric coefficient $g_{tt}\rightarrow -1$. 

According to traversable wormhole physics, the average null energy condition (NEC) is violation at the throat, and the method is check the Raychaudhuri equation. For general case, this equation reduce to $\rho+P_{r}\geq 0$. Based on the form of energy-momentum tensor $T_{\mu\nu} =diag(\rho,P_{r},P_{\theta},P_{\phi})$, the energy density $\rho$ and pressure $P_{r}$ are given by
\begin{equation}
\rho=\dfrac{1}{\Psi}-\dfrac{a^{2}(20y^{2}(k+a^{2})+24y^{2}F+(1-y^{2})k^{2}_{'r})}{4\Psi\Sigma^{4}}+$$$$\dfrac{3\Delta(H_{'r}+a^{2}H\psi_{'r})^{2}-4a^{4}y^{2}(1-y^{2})H^{2}\psi^{2}_{'r^{2}}}{4H^{2}\Psi}
+\dfrac{2a^{2}}{\Psi\Sigma^{2}}$$$$
+\dfrac{2a^{2}(a^{2}y^{2}(1+y^{2})-(1-3y^{2})k)\psi_{'y^{2}}}{\Psi\Sigma^{2}}-$$$$
\dfrac{1}{2H\Psi}\Big(8a^{2}y^{2}(1-y^{2})H\psi_{'y^{2}y^{2}}
+\Delta_{'r}(H_{'r}+a^{2}H\psi_{'r})+$$$$
2\Delta(H_{'rr}+a^{2}(2H_{'r}\psi_{'r}+H(a^{2}\psi^{2}_{'r}+\psi_{'rr})))\Big)
\label{EC1}
\end{equation}
and
\begin{equation}
P_{r}=-\rho+\dfrac{2a^{2}y^{2}\Delta}{\Psi\Sigma^{4}}-\dfrac{\Delta(H_{'r}k_{'r}+a^{2}Hk_{'r}\psi_{'r})}{H\Psi\Sigma^{2}}+$$$$
\dfrac{\Delta}{2H^{2}\Psi}\Big(3H^{2}_{'r}-2HH_{'rr}
+2a^{2}HH_{'r}\psi_{'r}+a^{4}H^{2}\psi^{2}_{'r}-2a^{2}H^{2}\psi_{'rr}\Big),
\label{EC2}
\end{equation}
where $\Psi=r^{2}+m^{2}+a^{2}\cos^{2}\theta$, $a^{2}\psi=ln[(k+a^{2}y^{2})/H]$, $k=H=r^{2}+m^{2}$ and $F=f(r)=1-\dfrac{2Mr^{k}}{\Big(r^{2n}+m^{2n}\Big)^{\frac{k+1}{2n}}}$. Therefore, for spacetime metric (\ref{Kerr-like1}), the NEC can be express as
\begin{equation}
P_{r}+\rho=\dfrac{2a^{2}y^{2}\Delta}{\Psi\Sigma^{4}}-\dfrac{\Delta(H_{'r}k_{'r}+a^{2}Hk_{'r}\psi_{'r})}{H\Psi\Sigma^{2}}+$$$$
\dfrac{\Delta}{2H^{2}\Psi}\Big(3H^{2}_{'r}-2HH_{'rr}
+2a^{2}HH_{'r}\psi_{'r}+a^{4}H^{2}\psi^{2}_{'r}-2a^{2}H^{2}\psi_{'rr}\Big).
\label{EC3}
\end{equation}
By calculation, we found that NEC was not satisfied even with the throats of wormholes, suggesting that these wormholes are just like Eills wormholes and Morris-Thorne wprmholes. Even so, this kind of wormhole is very interesting, because the value of $n$ and $k$ are different, the properties of wormhole will be greatly different.

\subsection{Quantum deformed black hole}
In the previous section, we analyzed the basic properties of the Kerr-like black-bounce metric. Here we will analyze the main properties of quantum deformation of black holes, including the structure of black hole event horizon, stationary limit surfaces, black hole singularity, non-zero components of energy-momentum tensors and energy conditions, etc. 

Firstly, from the equation (\ref{Kerr-like2}), the BH event horizon are determined by radial component, and satisfy condition
\begin{equation}
\Delta=r_{\pm}^{2}\Big((1-\dfrac{A^{2}}{r_{\pm}^{2}})^{\frac{n}{2}}-\dfrac{2M}{r_{\pm}}\Big)+a^{2}=0,
\label{QBH1}
\end{equation}
where $r_{+}$ and $r_{-}$ are EH and Cauchy horizon, respectivly. CaseI: $n=0$, the quantum deformation of BH reduce to Kerr BH, the two horizon are $r_{\pm}=M\pm \sqrt{M^{2}-a^{2}}$; CaseII: $n=1$, the spacetime is metric-regular, $r_{\pm}\sqrt{r_{\pm}^{2}-A^{2}}-2Mr_{\pm}+a^{2}=0$, the $r_{\pm}=\dfrac{1}{2}(A^{2}+4M^{2}\pm \sqrt{A^{4}+8A^{2}M^{2}+16M^{4}-4a^{4}})$; CaseIII: $n=2$, metric-regular, $r_{\pm}^{2}-2Mr_{\pm}+a^{2}-A^{2}=0$ lead to $r_{\pm}=M\pm\sqrt{M^{2}-a^{2}+A^{2}}$. 
A brief analysis is made here, and the contribution of quantum fluctuation to the BH is completely reflected in $A$. It can be seen that the effect of $A$ is opposite to the BH spin $a$. The increase of the BH spin reduces the radius of the BH event horizon, while the increase of the quantum fluctuation parameter increases the radius of the BH event horizon. 

Secondly, the SIS of quantum deformed BH are determined by $g_{tt}=0$, which reduce to
\begin{equation}
1-\dfrac{r_{\rm{SLS\pm}}^{2}+2Mr_{\rm{SLS\pm}}-r_{\rm{SLS\pm}}^{2}\Big(1-\dfrac{A^{2}}{r_{\rm{SLS\pm}}^{2}}\Big)^{\frac{n}{2}}}{r_{\rm{SLS\pm}}^{2}+a^{2}\cos^{2}\theta}=0,
\label{QBH1}
\end{equation}
where $r_{\rm{SLS+}}$ and $r_{\rm{SLS-}}$ are out SIS and inner SIS, respectivly. CaseI: $n=0$, the quantum deformation of BH reduce to Kerr BH, the two SIS are $r_{\rm{SLS\pm}}=M\pm \sqrt{M^{2}-a^{2}\cos^{2}\theta}$; CaseII: $n=1$, the spacetime is metric-regular, $r_{\rm{SLS\pm}}\sqrt{r_{\rm{SLS\pm}}^{2}-A^{2}}-2Mr_{\rm{SLS\pm}}+a^{2}\cos^{2}\theta=0$, the $r_{\rm{SLS\pm}}=\dfrac{1}{2}(A^{2}+4M^{2}\pm \sqrt{A^{4}+8A^{2}M^{2}+16M^{4}-4a^{4}\cos^{4}\theta})$; CaseIII: $n=2$, metric-regular, $r_{\rm{SLS\pm}}^{2}-2Mr_{\rm{SLS\pm}}+a^{2}\cos^{2}\theta-A^{2}=0$ lead to $r_{\rm{SLS\pm}}=M\pm\sqrt{M^{2}-a^{2}\cos^{2}\theta+A^{2}}$. 
The generation of quantum fluctuations reduces SIS, which reduces the ergosphere region of the rotating BH. It can be speculated that such an effect would have an impact on the Blandford-Zinjek mechanism, particle acceleration and so on.

Thirdly, by calculating the Kretsmann scalar $R$ corresponding to the BH, it can be found that the singularity structure of the BH is exactly the same as that of Kerr BH, and the quantum correction parameter $A$ does not have any effect on the singularity, which is very difficult to understand.

\section{Summary}
\label{sum}
In this work we obtain two exact solutions of Einstein$'$s field equation, the rotating spacetime of black-bounce and quantum deformed BH. The properties of these rotating spacetimes are discussed in detail and some interesting properties are found. For rotating black-bounce spacetime, the BH spin changes the value of critical parameter $m_{c}$, and found that it satisfies the following range $\sqrt{2}M\leqslant m_{c}\leqslant2M$. This is very different from the spherical symmetric case. For rotating quantum deformed BH, we find that the effect of quantum modification parameters on BHs is opposite to the effect of BH spin, including black hole event horizon and SIS, which may indicate some repulsive effect of quantum effects on space-time. It is hard to understand that the singularity of BHs is not affected by quantum corrections.

\begin{acknowledgments}
We acknowledge the anonymous referee for a constructive report that has significantly improved this paper. We acknowledge the  Special Natural Science Fund of Guizhou University (grant
No. X2020068) and the financial support from the China Postdoctoral Science Foundation funded project under grants No. 2019M650846.
\end{acknowledgments}

\end{document}